\documentclass[aps,prd,english,showpacs,11pt]{revtex4-1}
\usepackage[latin1]{inputenc}
\usepackage{ucs}
\usepackage{amsmath}
\usepackage{amsfonts}
\usepackage{amssymb}
\usepackage{graphicx}
\usepackage{subcaption}
\usepackage{psfrag}
\usepackage{picinpar}
\usepackage[]{hyperref}

\begin{document}
\title{Superfluid to normal fluid phase transition in the Bose gas trapped in two dimensional optical lattices at finite temperature}

\author{M. O. C. Pires}
\email{marcelo.pires@ufabc.edu.br}
\affiliation{Centro de Ci\^{e}ncias Naturais e Humanas, Universidade
  Federal do ABC, Rua Santa Ad\'elia 166, 09210-170, Santo Andr\'{e}, SP,
  Brazil}

\author{E. J. V. de Passos} 
\email{passos@fma.if.usp.br}
\affiliation{Instituto de F\'{i}sica, Universidade de S\~ao Paulo, CP 66318,
  05315-970, S\~ao Paulo-SP, Brazil}

\begin{abstract}
We develop the Hartree-Fock-Bogoliubov theory at finite temperature for Bose gas trapped in the two dimensional optical lattices. The on-site energy is considered low enough that the gas presents superfluid properties. We obtain the condensate density as function of the temperature neglecting the anomalous density in the thermodynamics equations. The condensate fraction provide two critical temperature. Below the temperature $T_{C1}$ there is one condensate fraction. Above two possible fractions merger up to the  critical temperature $T_{C2}$. Then the gas provides an first order transition at temperature above $T_{C2}$ where the condensate fraction is null. We resume by a finite-temperature phase diagram where can be identify three domains: the normal fluid, the superfluid and the superfluid with two possible condensate fractions. 
\end{abstract}


\maketitle

\section{Introduction}

The Bose-Einstein condensate trapped in the
periodic potential provided a peculiar phenomenum in which there exist the quantum phase transition in the ultra-cold atoms similar to the phase transition seen in crystals \cite{mor06,blo08}.  In this transition, the variation of external parameter related to the trap potential modifies the non-local superfluid properties into the local properties of the Mott insulator phase.

At low temperatures, the superfluid-to-Mott-insulator (SF-MI) quantum
phase transition was proposed in 1998 by Jaksch et
al. \cite{jak98} for ultra-cold gases. They considered a Bose-Hubbard model to describe the Bose gas on the optical lattices, and they showed the quantum phase transition by changing the on-site energy, the hopping matrix element or the chemical potential.
Four years later, Greiner et
al. \cite{gre02} detected the SF-MI phase transition by imaging atoms
released from optical lattices with different periodic
potentials. Above a critical on-site energy, the atomic distribution
presents the correlation function concerned to the Mott-insulator.  In
this phase, the position of atoms \cite{gem09} become a tool to probe
and measure thermodynamic properties such as pressure, temperature,
and transport properties \cite{hun10}.

Another way to observe the SF-MI phase transition is analysing the
collective excitations in the gas trapped in optical lattices. The
low-lying excitation energy of the gas is achieved in 1D via lattice
modulation \cite{sto04} or by Bragg spectroscopy \cite{cle09}. In the
superfluid (SF) phase, the energy spectrum presents a gapless behaviour
which can be explained within a Bogoliubov theory \cite{oos01}, or in
terms of path integral formalism \cite{kle12} or by Green's function
formalism \cite{zal12}. 


In Mott insulator (MI) phase, the gapped excitation mode is  found
by the mean-field theory \cite{oos01} using the particle-hole
excitations frame of the superfluity modes. The mean-field theory in
\cite{oos01} neglected the quantum fluctuations, then the description of the system near to SF-MI phase transition is unavailable and, therefore, not confirmed the gapped mode \cite{bis11}. The SF-MI transition is
analysed in terms of a strong-coupling perturbation theory which
treats the kinetic energy as a mean-field and deals exactly the
on-site repulsion \cite{fis89,jak98}. This approximation gave a estimative for the SF-MI transition limits. 


The thermal gapless properties of ultra-cold bosons in
optical lattices were discussed by Gerbier in the
ref. \cite{ger07}. However, his analysis were done considering only the Mott insulator phase. At superfluid phase, a finite-temperature study of ultra-cold bosons in 2D optical lattices were done in \cite{mah12} by quantum
Monte-Carlo simulation taking account the effects of harmonic confinement. In
this reference, they observed the spatial inhomogeneity allows the
coexistence of three phases: the usual ground state superfluid, the Mott insulator phase
and the normal fluid phase. Their simulations can be compared with the phase diagram provided by the experimental results \cite{jim10}.

The phase diagram is describe as a function of the temperature and the interaction strength whose definition is the ratio between the on-site interaction strength  and the hopping matrix element. The phase diagram for the homogeneous 3D system at unity filling is plotted using the quantum Monte Carlo simulation in the reference \cite{cap07} and shown experimentally in \cite{tro10}. At zero temperature, the system undergoes a SF-MI quantum phase transition at the critical interaction strength. At temperatures different from zero and near to the critical interaction strength, they observed a suppression of the critical temperature where the condensate fraction is null and all particles are in the thermal cloud.

In this work, we investigate the thermal behaviour of the ultra-cold
Bose gas in 2D optical lattices on the superfluid phase and far from the SF-MI phase transition by a finite-temperature mean-field theory. This theory is known as Hartree-Fock-Bogoliubov (HFB) theory and departing from the
Bose-Hubbard model. Although the HFB theory provide a gapped spectrum
even far from the quantum phase transition, we avoid this problem by the
Popov approximation where the anomalous densities are neglected {\it
  ad hoc}. Hence the spectrum is gapless and, as theoretically shown
in \cite{oos01} and experimentally observed in \cite{bis11}, the
condensate fraction has the superfluid properties and depends on the
temperature and the interaction strength. Then, in
HFB theory, the superfluid phase has, as order parameter, the fraction of
condensate. We determine the order parameter as function of the temperature at a regime below to the quantum phase transition.


In sec. \ref{hfbsec} we obtain the HFB theory considering the
superfluid state as a vacuum of the quasi-particles at finite
temperature, we determine the limits to the Bogoliubov theory and we
implement the Popov approximation to derive the condensate
fraction. The numerical solution for the condensate fraction in the
Popov approximation is evaluated in the sec. \ref{numerical}. Finally,
in sec. \ref{conclusion}, we discuss the properties of the system near
the thermal phase transition according to the solution of the
mean-field theory.

\section{Hartree-Fock-Bogoliubov approximation at finite temperature}
\label{hfbsec}

The dynamics of the 2D ultra-cold boson gas in the optical lattices can be done by the second quantized Hamiltonian,
\begin{eqnarray}
H&=&\int\psi^\dagger({\bf
  x})\left(-\frac{\hbar^2\nabla^2}{2m}+V_0({\bf x})+V_T({\bf
  x})\right)\psi({\bf x})d^3{\bf x}+ \\ \nonumber
&+&\frac{1}{2}\frac{4\pi a_s\hbar^2}{m}\int\psi^\dagger({\bf
  x})\psi^\dagger({\bf x})\psi({\bf x})\psi({\bf x})d^3{\bf x}
\end{eqnarray}
where $\psi({\bf x})$ is the boson field operator corresponding to the
atom with the mass equal to $m$, $V_T({\bf
  x})=\frac{m}{2}\sum_{j=1}^{3}\omega_j^2x_j^2$ is the
magneto-optical trap and $V_0({\bf x})=V_{0}(\sin^2(\pi x/a)+\sin^2(\pi y/a))$ is the external
periodic potential formed by the intersecting laser beams with $V_0$
being the trap depth proportional to the laser intensity, $a$ being the
less distance between two sites. In this Hamiltonian, the inter-atomic interaction corresponds to the low-energy collision characterized by the $s$-wave scattering length, $a_s$.

We expand the field operators in Wannier functions
whose asymptotic behaviour is adjusted by the Gaussian function, $w({\bf
  x})=e^{-(x^2+y^2)/2\sigma^2}/(\pi^{1/2}\sigma)^{1/2}$
with $\sigma=\hbar/(m^2V_0E_R)^{1/4}$ and
$E_R=\hbar^2\pi^2/(2ma^2)$ is the recoil energy. Considering only the
lowest vibrational state in each site, the field operator can be written as, $\psi({\bf x})=\sum_iw({\bf x}-{\bf x}_i)\alpha_i$,
where ${\bf x}_i$ is the position of the site labelled by $i$ in the
2D lattices with $N_s$ sites. Thus, in terms of the Wannier operators
$\alpha_i$, the Hamiltonian become a Bose-Hubbard (B-H)
Hamiltonian,
\begin{eqnarray}
H=-t\sum_{\langle i,j\rangle}\alpha_i^\dagger
\alpha_j+\frac{1}{2}U\sum_i\alpha_i^\dagger \alpha_i^\dagger \alpha_i
\alpha_i,
\end{eqnarray}
where $t=-\int w^*({\bf
  x}-{\bf x}_i)\left(-\frac{\hbar^2\nabla^2}{2m}+V_0({\bf x})\right)w({\bf x}-{\bf x}_j)d^3{\bf x}$ is the hopping parameter and
$U=\frac{4\pi a_s \hbar^2}{m}\int |w({\bf x})|^4d^d{\bf x}$ is the
on-site interaction strength. The hopping parameter is the same for any site due to neglect the magneto-optical trap effects, $V_T({\bf x})=0$.

In order to identify the condensate state, we expand the Wannier operators
in the 2D quasi-momentum space,
$\alpha_i=\frac{1}{\sqrt{N_s}}\sum_{\bf k}a_{\bf k}e^{-\imath(k_xx_i+k_yy_i)}$, where the component of quasi-momentum, ${\bf k}$,
assume discrete values give by $k_j=2\pi\eta/(M_ja)$ with
$\eta=0,\pm1,\pm2,\cdots,\pm(M_j-1)/2$ and $M_j$ is the even number corresponding to the number of site in the $j$-direction.

The condensate state is a coherent state of $N_0$ single particles with quasi-momentum equal to zero. We represent this state in the B-H Hamiltonian by shifting the zero quasi-momentum operator, $a_{\bf
  k}=b_{\bf k}+\sqrt{N_0}\delta_{\bf 0,k}$. Then we have the shifted B-H Hamiltonian as,
\begin{eqnarray}
H&=&(-\bar{\epsilon}_0-\mu)N_0+\frac{g}{2}N_0^2+\sqrt{N_0}(-\bar{\epsilon}_0-\mu+gN_0)(b_0+b^\dagger_0)+\\ \nonumber
&+&\sum_{\bf k}\left[(-\bar{\epsilon}_{\bf k}-\mu+2gN_0)b^\dagger_{\bf k}b_{\bf k}+\frac{g}{2}N_0(b_{\bf k}b_{-\bf k}+b^\dagger_{\bf k}b^\dagger_{-\bf k})\right]+ \\ \nonumber
&+&z_0g\sum_{\bf kq}(b^\dagger_{\bf k+q}b^\dagger_{\bf k}b_{\bf q}+b^\dagger_{\bf k-q}b_{\bf k}b_{\bf q})+\frac{g}{2}\sum_{\bf kk'q}b^\dagger_{\bf k+q}b^\dagger_{\bf k'-q}b_{\bf k}b_{\bf k'},
\label{bhh}
\end{eqnarray}
where $\bar{\epsilon}_{\bf k}=2t(\cos(k_x a)+\cos(k_y a))$ is the energy of the free quasi-particles, $g=U/N_s$ is the interaction
parameter and $\mu$ is the chemical potential in which ensures a
well-defined particle number even for the particle reservoir contact.

For the purpose of decoupling normal modes, we introduce the Bogoliubov transformation, $b^\dagger_{\bf k}=u_{\bf
  k}c^\dagger_{\bf k}+v_{\bf k}c_{\bf -k}$ where $u_{\bf k}$ and
$v_{\bf k}$ are the real Bogoliubov coefficients and obey $u^2_{\bf
  k}-v^2_{\bf k}=1$ to keep the commutation relations. The Bogoliubov coefficients are adjusted for becoming the B-H Hamiltonian in diagonal form, $H=E_0+\sum_{\bf k}E_{k}c^\dagger_{\bf k}c_{\bf k}$, where $E_0$
is the ground state energy and $E_{k}$ is the excitation energy.

Assuming the time invariance of the Bogoliubov coefficients, we derive the excitation energy as, $
E_{k}=\sqrt{h^2_{k}-\Delta^2}$,
and the Bogoliubov coefficients are, $v^2_{k}=\frac{1}{2}\left(\frac{h_{k}}{E_{k}}-1\right)$ and $u_{k}v_{k}=-\frac{1}{2}\frac{\Delta}{E_{k}}$,  
where the parameters $h_{k}=\bar{\epsilon}_0-\bar{\epsilon}_k+g(N_0-\sum_{k}\kappa_k)$ and $\Delta=g(N_0+\sum_{k}\kappa_k)$ depend on the variational parameters, 
$\rho_{k}=\nu(k,T)+(1+2\nu(k,T))v^2_{k}$ and $\kappa_{k}=(1+2\nu(k,T))u_{k}v_{k}$. These parameters are thermal averages of shifted operator and depend on the temperature \cite{bla87}. The function $\nu(k,T)=(e^{E_k/k_BT}-1)^{-1}$ is the Bose-Einstein distribution of a quantum particle with energy equal to $E_k$. 

Finally, we can obtain the chemical potential in terms of the number of the condensed particles, $N_0$, the depleted particles, $\sum_{k}\rho_k$ and the paring, $\sum_{k}\kappa_k$. The chemical potential is written as $
\mu=-\bar{\epsilon}_0+g[N_0+\sum_{k}(2\rho_k+\kappa_k)]$.

\subsection{Particle number}

To evaluate the condensate fraction $n_0=N_0/N_s$ as an explicit function of the dimensionless interaction strength, $U/t$, and the temperature, $T$, we introduce the total density in each site by, $n=\sum_{\bf k}\langle a^\dagger_{\bf k}a_{\bf k} \rangle/N_s$. In terms of the Bogoliubov parameters, the depletion density is given by, $n-n_0=\sum_{\bf k}\rho_{\bf k}/N_s$. Hence,
we have the condensate density in each site as,
\begin{eqnarray}
n_0=n-\frac{1}{N_s}\sum_{\bf
  k}\frac{1}{2}\left(\coth\left(\frac{E_k}{2k_BT}\right)\frac{h_{\bf
    k}}{E_{k}}-1\right),
\end{eqnarray}
where, $h_{\bf k}=\zeta_k+U(n_0-m_0)$, $\Delta=U(n_0+m_0)$ and
$\zeta_{k}=4t[\sin^2(k_xa/2)+\sin^2(k_ya/2)]$.

The anomalous density in each site, $m_0=(\sum_k\kappa_k)/N_s$, can be an explicit form that depend on the condensate density,
$m_0=-n_0(\sum_k\coth(E_k/2k_BT)/E_k)/(2/U+\sum_k\coth(E_k/2k_BT)/E_k)$.
If the anomalous density is different from zero, we have a gap in the excitation spectrum given by,
$\lim_{{k}\to 0}E_k=2U(-n_0m_0)^{1/2}$.


\subsection{Popov approximation}

In the HFB theory, the excitation spectrum presents a gap for long-wavelength excitations and, therefore, not satisfies the
Hugenholtz-Pines theorem \cite{gri96,hoh65}. Indeed, the gap in the
spectrum for long-wavelength can not explain the superfluid
properties, as the absence of viscosity \cite{lan80,gri93}, widely
identified due to the phonon behaviour of the excitation spectrum in
ultra-cold alkaline atoms \cite{ket00}.

At very low temperatures and in the very weakly interaction regime, the condensate fraction is high enough to neglect the depletion, $n_0\approx n$, and the anomalous density $m_0=0$. By this approximation, known as Bogoliubov approximation, the excitation energy, $E_k=\sqrt{\zeta_{k}(\zeta_{k}+2nU)}$, is gapless for the long-wavelength limit and the condensate density
depends only on the total density, $n$.

Otherwise, we can neglect
{\it ad hoc} only the anomalous density, $m_0=0$. The so-called Popov approximation provides  a non-linear integral equation to the condensate density given by,
\begin{eqnarray}
n_0=n-\frac{1}{2N_s}\sum_{\bf
  k}\frac{\coth\left(E_k/2k_BT\right)(\zeta_{k}+n_0U)-E_k}{E_k},
\label{cd}
\end{eqnarray}
with a gapless spectrum equal to $E_k=\sqrt{\zeta_{k}(\zeta_{k}+2n_0U)}$.

\section{Numerical results}
\label{numerical}

We obtain the condensate fraction, $n_0/n$, as function of the temperature and the dimensionless interaction strength parameter, $U/t$,  solving numerically the
non-linear equation (\ref{cd}) derived by the Popov approximation.



\begin{figure}[ht]
  \begin{center}
    \includegraphics[width=.85\textwidth]{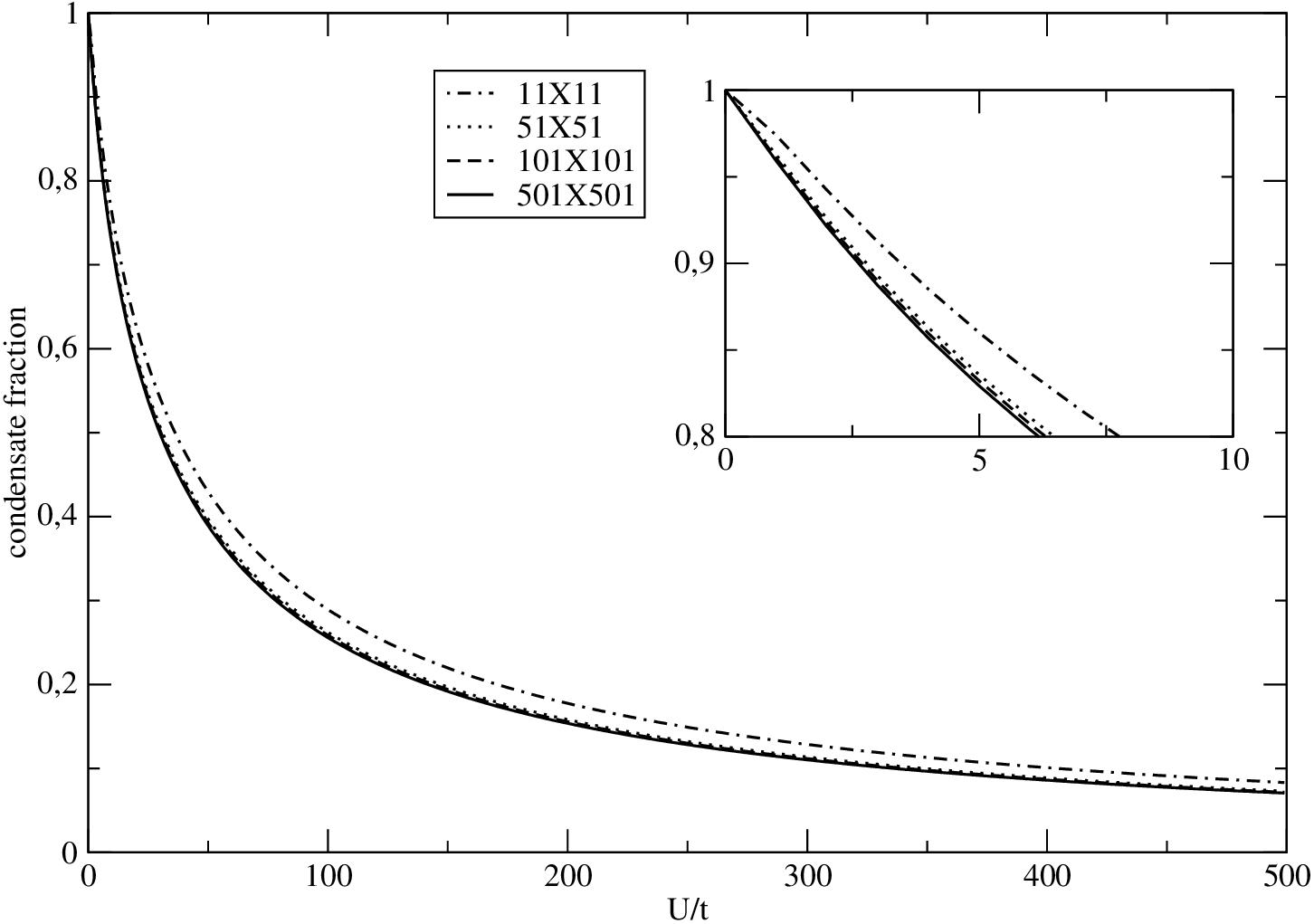}
    \caption{\footnotesize The condensate fraction $n_0/n$ in
      two-dimensional lattice as function of the dimensionless
interaction strength parameter $U/t$ at zero temperature. The graph exhibits the
      approach to the thermodynamic limit in which the ration
      $N/N_s=n$ remains constant.  This approach is realized when the
      size of the lattice exceeds $10^{4}$ sites.}
 \label{plot03} 
 \end{center}
\end{figure}

At zero temperature, we plot in fig. \ref{plot03} the condensate fraction as function of the interaction strength for four sizes of the 2D squared lattice. When we increase the size of the lattice, the condensate fraction converge to an asymptotic curve regarded to the thermodynamic limit.  Then, as the difference between
the curves provide for lattice with $101\times 101$ and $501\times 501$ is small and for the sake of the computation time, we assume the size of lattice equal to $101\times 101$ can be considered in
the thermodynamic limit. 

In the ref. \cite{oos01}, they analysed the limits for the strongly interacting regime, $U/t\to\infty$, at zero temperature and they concluded that the quantum phase transition could not be identified in the Bogoliubov approximation. However the arguments used to justify the absence of phase transition are based on the Popov approximation indeed.
We observe the same result in the fig. \ref{plot03} for any size of lattice using the Popov approximation. Differently from them,  it is not clear for us the rule of the Bogoliubov approximation in this lack of the quantum phase transition.

\vskip 4em

\begin{figure}
  \begin{center}
    \includegraphics[width=.85\textwidth]{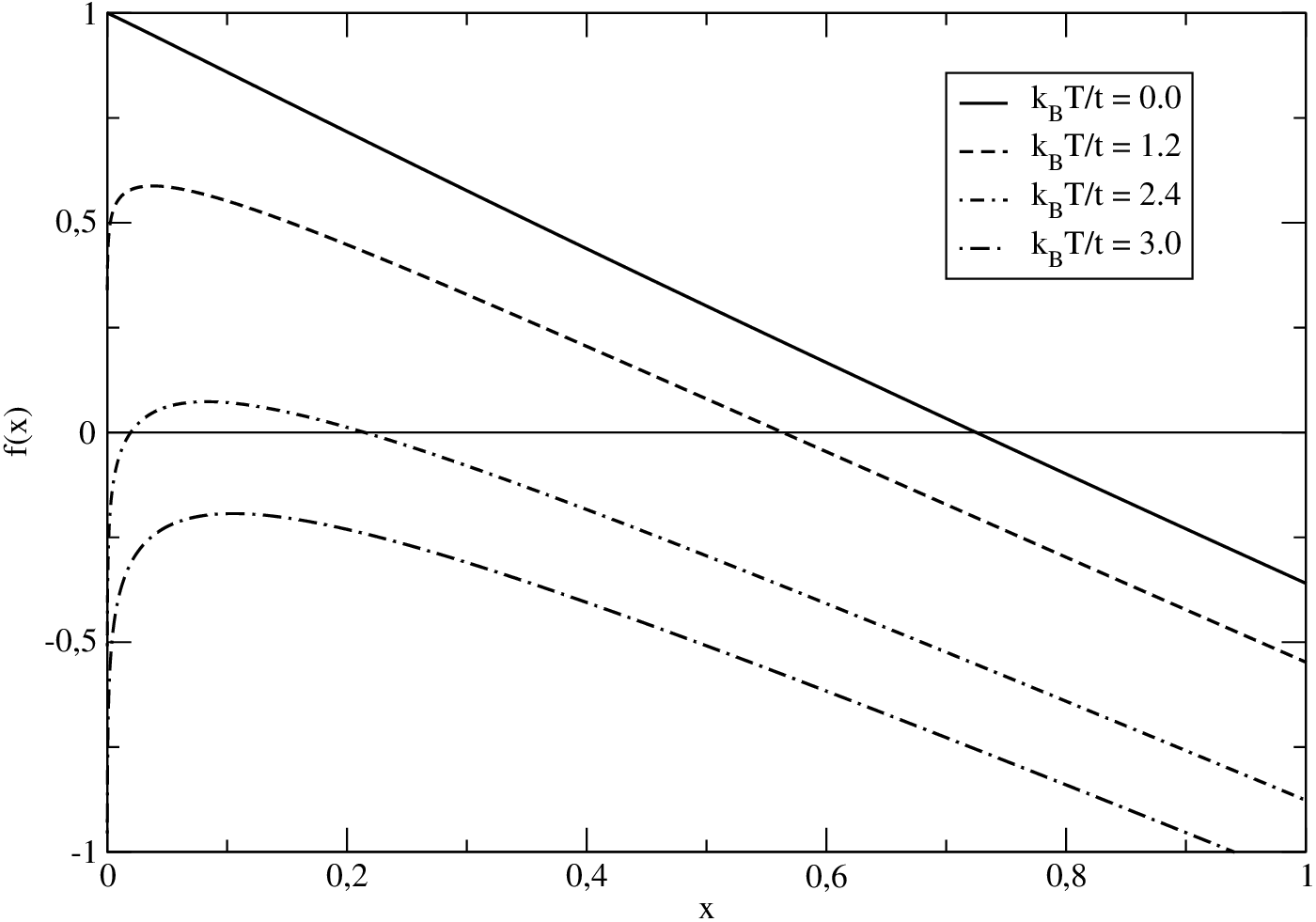}
    \caption{\footnotesize The function $f(x)$ (eq. \ref{lead}) is
      plotted for different values of the temperatures and for $n=1.0$,
      $U/t=10.0$ and $N_S=101\times 101$. This graph illustrates the
      numerical method used to find the roots of $f(x)$.}
%
%
 \label{f0} 
 \end{center}
\end{figure}

In order to obtain the temperature-dependent condensate fraction, we introduce the function,
\begin{eqnarray}
f(x)=1-x-\frac{1}{nN_s}\sum_{\bf
  k}\frac{1}{2}\left(\coth\left(\frac{\sqrt{\epsilon_{\bf
      k}(\epsilon_{\bf k}+2nUx)}}{2k_bT}\right)\frac{\epsilon_{\bf
    k}+nUx}{\sqrt{\epsilon_{\bf k}(\epsilon_{\bf k}+2nUx)}}-1\right),
\label{lead}
\end{eqnarray}
where the variable $x$ running from $0$ to $1$. The function $f(x)$ is
determined by the temperature $T$ and the interaction strength parameter
$U/t$ and its roots are the condensate fractions at this temperature and for this interaction strength. In fig. \ref{f0} we plot this function for $U/t=10.0$ and at different temperatures. 
The plot exhibits three different types of the
solution for equation (\ref{cd}). There is one solution for $0\le T\le
T_{C1}$, where $T_{C1}$ is the temperature in which the function
$f(x)$ presents a root at $x=0$. When we increase the temperature, we
find two solution which merger at $T=T_{C2}$. At temperatures larger than $T_{C2}$, there is no solution and, therefore, there is no condensate fraction in the lattice.

\vskip 2em

\begin{figure}
  \begin{center}
    \includegraphics[width=.85\textwidth]{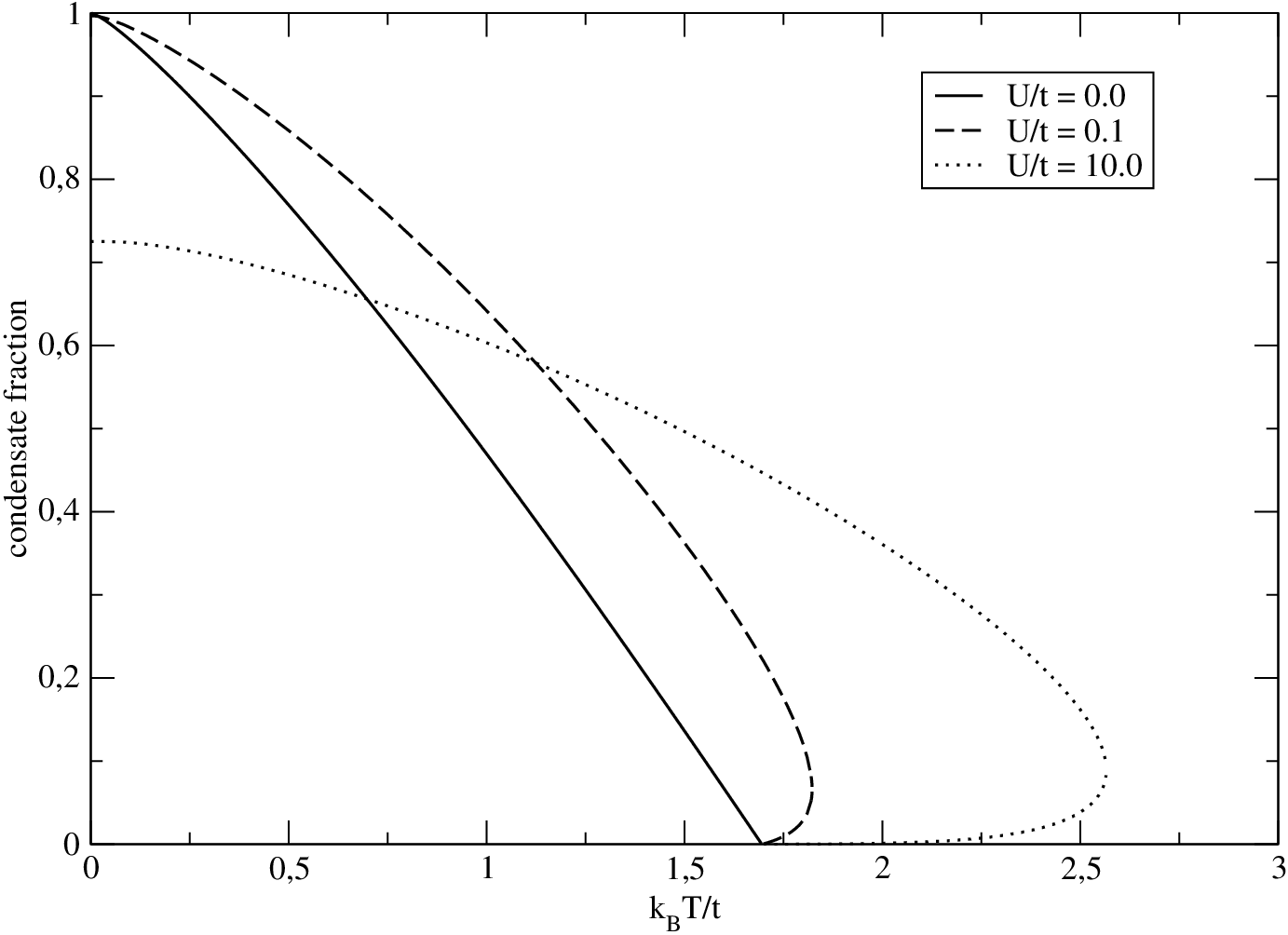}
    \caption{\footnotesize The condensate fraction $n_0/n$ as a
      function of the temperature $T$ for different values of the
      interaction strength parameter $U/t$ in the same conditions of
       \ref{f0}. Three interacting regimes are deal in the figures. The
      non-interacting gas is describe by the solid curve, the weakly
      interacting gas is shown in the traced line, and the strongly
      interacting gas is represented by the dotted curve.}
%
 \label{f1} 
 \end{center}
\end{figure}

In the fig. \ref{f1} we plot the condensate fraction as function of
the temperature for different values of the interaction strength. In the non-interacting case $U/t=0.0$, the fraction
decrease when the temperature increase and reach a null fraction,
i. e., the condensate disappear at the temperature $T_{C1}=1.6954$ (in
units of $k_B/t$). Above this critical temperature the
gas is entirely a thermal Bose fluid and the quasi-particles have free
energy dispersion in the limit of the large wavelength.

For the interacting case $U/t\ne 0.0$, there are only one condensate
fraction in the interval of the temperature $0\le T\le
T_{C1}$. In this condition, the values of the fraction decrease
as temperature increase. For the temperature in the range of
$T_{C1}<T<T_{C2}$, there are two possible condensate
fractions. Differently from the largest fraction which decreases as the
temperature increases, the smallest fraction increases up to merger the
highest fraction at the second critical temperature $T_{C2}$. At
temperature above $T_{C2}$, there is no solution for $f(x)$ and,
therefore, there is no condensate in the lattice.

The fig. \ref{f1} shows that exist an interval of the temperature in
which the strongly interacting Bose gas ($U/t=10.0$) presents the
condensate fraction larger than the condensate fraction of the weakly
interacting Bose gas ($U/t=0.1$). This unusual behaviour is contrary
to what intuition or common sense would indicate.

\vskip 2em  

\begin{figure}
  \begin{center}
    \includegraphics[width=.85\textwidth]{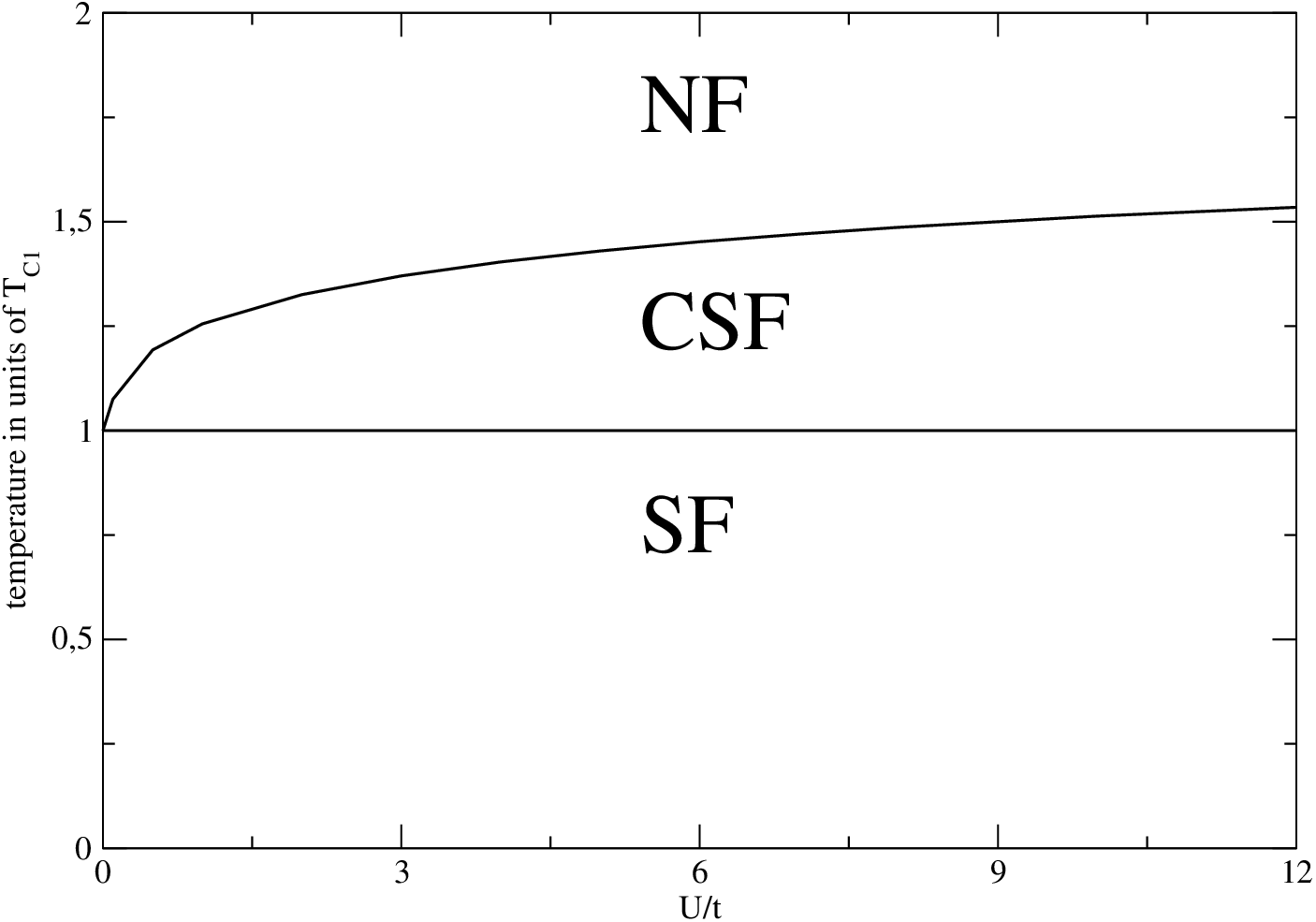}
    \caption{\footnotesize Finite-temperature phase diagram as a function of dimensionless parameter $U/t$ and temperature $T$ of the Bose gas
      trapped in two dimensional lattice with size of $N_S=101\times
      101$ and density equal to $n=1.0$.}
%
 \label{f2} 
 \end{center}
\end{figure}

The phase diagram shown in the fig. \ref{f2} has three regions
correspond to the possible values of the condensate fraction. The
region SF where there is one condensate fraction, CSF in which
coexists two possible condensate density and the region NF composed by
the thermal free quasi-particles. As we increase the interaction strength parameter, the CSF region increases monotonically.

\begin{figure}
  \begin{center}
    \includegraphics[width=.85\textwidth]{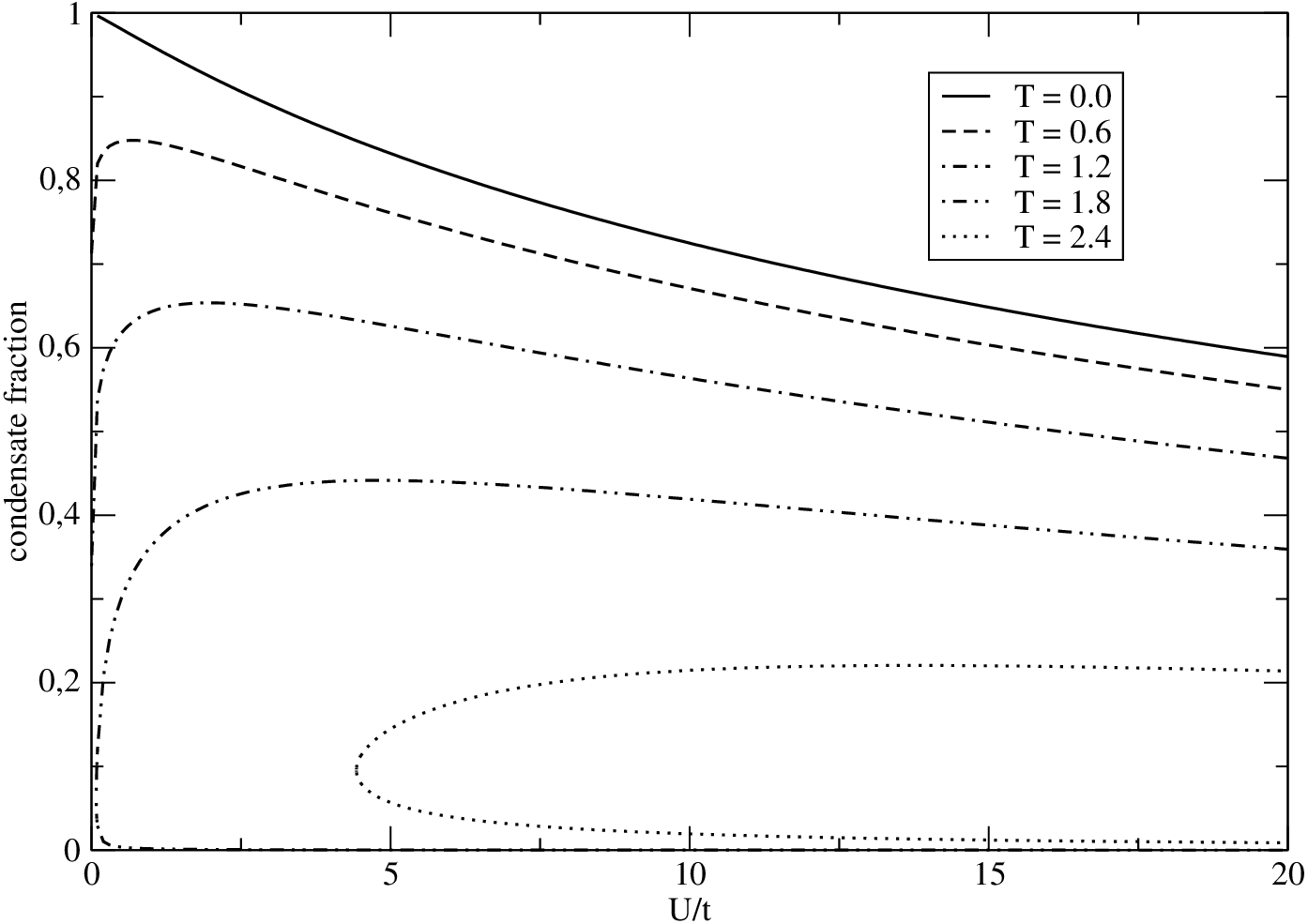}
    \caption{\footnotesize The condensate fraction $n_0/n$ as a
      function of the dimensionless parameter $U/t$ at different
      temperatures (in units of $t/k_B$) with $n=1.0$ and
      $N_S=101\times 101$. At temperatures below $T_{C1}$, the curves
      are positive functions with only one maximum point. At
      temperature above, the curves present a minimum critical value
      for the dimensionless parameter $U/t$ where below there is no
      condensate at this temperature. For above the critical
      dimensionless parameter, the condensate can coexist in two
      possible condensate fraction.}
 \label{plot05} 
 \end{center}
\end{figure}

Finally, in the fig. \ref{plot05} we present the condensate fraction
$n_0/n$ in two dimensional lattice as a function of the interaction strength parameter at several temperatures. At temperature
below the critical temperature $T_{C1}$, the condensate fraction
increase when the interaction parameter increase. This growth is
provided by the positive contribution of the interaction. The
repulsive interaction promotes the non-localization of the atoms in
the lattice. Although the effect of the interaction, the entropy plays
the leading role for large values of the interaction parameter. Then, the maximum value of the condensate fraction is
reached around $U/t\approx 0.5$, and the fraction decreases
monotonically up to zero in the limit of strongly interaction. At
temperatures higher than the critical temperature $T_{C2}$, the curves
present a minimum value to the interaction parameter in which the
two possible solution of the condensate fraction rise. As the case of
low-temperature, there is a balance between the interaction and the
entropy. For interaction parameter close to the critical parameter,
the role of the interaction is more determinant than the
entropy. However the entropy dominates for the limit of the strongly interacting regime.

\section{Summary}
\label{conclusion}

We develop the Hartree-Fock-Bogoliubov theory at finite temperature
for Bose gas trapped in the two dimensional optical lattices. This
mean-field theory presents anomalous pairing density in which provides a gap in the energy spectra for large wavelength excitation.

To obtain a gapless excitation energy, we consider the Popov
approximation where the anomalous density is neglected {\it ad hoc}. In
the Popov approximation, we derive a non-linear equation whose
solutions give us the condensate fraction in the system.  Considering
the size of the lattices large enough to be in the thermodynamic
limit, we obtain curve of condensate fraction as function of the
temperature. These curves present two critical temperature, one of them
$T_{C1}$ define an interval where we have a single order parameter. Above
this temperature, there are two condensate fraction that merger at
the critical temperature $T_{C2}$. Then the gas provides an first order
transition when temperature pass over $T_{C2}$ and, in this condition, the condensate fraction is null. We identify an interval of the temperature in which the
strongly interacting gas has the condensate fraction large than the
weakly interacting gas. This unusual effect is due to the balance
between the role of the entropy and the interaction. 

A phase diagram as a function of interaction strength $U/t$ and temperature $T$ is done calling attention to the regions where occur the normal fluid phase, the superfluid phase and the superfluid phase in which coexist two different condensate fractions. There is a decay behaviour when the interaction strength approach to the zero as observed by \cite{cap07} using the quantum Monte Carlo simulation for 3D lattice. Then the HFB theory is qualitatively usefull to describe the condensate fraction in the weakly interacting regime. The quatitative comparison between the HFB theory and the Monte-Carlo simulation is out of scope and it will be done in future work.  Otherwise, for large interaction strength, the thermal phase transition temperature increase and, therefore, frustrates any possibility to identify the quantum phase transition by the mean-field theory. Indeed, the trial function to develop the HFB theory is not appropriated to describe the strongly correlated lattice.

\begin{acknowledgments}
E. J. V. P.  thanks for CAPES for financial support.
\end{acknowledgments}


\begin{thebibliography}{References}

\bibitem{mor06} O. Morsch and M. Oberthaler, {\it Rev. Mod. Phys.}
  {\bf 78}, 179 (2006).

\bibitem{blo08} I. Bloch, J. Dalibard, and W. Zwerger, {\it
  Rev. Mod. Phys.} {\bf 80}, 885 (2008).

\bibitem{jak98} D. Jaksch, C. Bruder, J. I. Cirac, C. W. Gardiner, and
  P. Zoller, {\it Phys. Rev. Lett.} {\bf 81}, 3108 (1998).

\bibitem{gre02} M. Greiner, O. Mandel, T. Esslinger, T. W. Hansch, and
  I. Bloch, {\it Nature} {\bf 415}, 39 (2002).


\bibitem{gem09} N. Gemelke, X. Zhang, C. Hung, and C. Chin, {\it
  Nature} {\bf 460}, 995 (2009).

\bibitem{hun10} C. L. Hung, X. Zhang, N. Gemelke, and C. Chin, {\it
  Phys. Rev. Lett.} {\bf 104}, 160403 (2010).
  

\bibitem{sto04} T. Stoferle, H. Moritz, C. Schori, M. Kohl, and
  T. Esslinger, {\it Phys. Rev. Lett.} {\bf 92}, 130403 (2004).

\bibitem{cle09} D. Clement, N. Fabbri, L. Fallani, C. Fort, and
  M. Inguscio, {\it Phys. Rev. Lett.} {\bf 102}, 155301 (2009).

\bibitem{oos01} D. van Oosten, P. van der Straten, and H. T. C. Stoof,
  Phys. Rev. A {\bf 63}, 03601 (2001).

\bibitem{kle12} H. Kleinert, Z. Narzikulov, and A. Rakhimov, {\it
  Phys. Rev. A} {\bf 85}, 063602 (2012).

\bibitem{zal12} T. A. Zaleski, {\it Phys. Rev. A} {\bf 85}, 043611
  (2012).
 

\bibitem{bis11} U. Bissbort, Y. Li, S. Gotze, J. Heinze,
  J. S. Krauser, M. Weinberg, C. Becker, K. Sengstock, and
  W. Hofstetter, {\it Phys.  Rev. Lett.} {\bf 106}, 205303 (2011).

\bibitem{fis89} M. P. A. Fisher, P. B. Weichman, G. Grinstein, and
  D. S. Fisher, {\it Phys. Rev. B} {\bf 40}, 546 (1989).







\bibitem{ger07} F. Gerbier, {\it Phys. Rev. Lett.} {\bf 99}, 120405
  (2007).

\bibitem{mah12} K. W. Mahmud, E. N. Duchon, Y. Kato, N. Kawashima,
  R. T. Scalettar, and N. Trivedi, {\it Phys. Rev. B} {\bf 84}, 054302
  (2011).

\bibitem{jim10} K. Jimenez-Garcia, R. L. Compton, Y.-J. Lin,
  W. D. Phillips, J. V.Porto, and I. B. Spielman, {\it
    Phys. Rev. Lett.} {\bf 105}, 110401 (2010).

\bibitem{cap07} B. Capogrosso-Sansone, N. V. Prokof'ev, and B. V. Svistunov, {\it Phys. Rev. B} {\bf 75}, 134302 (2007).

\bibitem{tro10} S. Trotzky, L. Pollet, F. Gerbier, U. Schnorrberger, I. Bloch, N. V. Prokof'ev, B. Svistunov, and M. Troyer, {\it Nature Phys.} {\bf 6}, 998 (2010).

\bibitem{bla87} J. -P. Blaizot and G. Ripka, Quantum Theory of Finite
  Systems (MIT Press, Cambridge, 1987).

\bibitem{gri96} A. Griffin, Phys. Rev. B {\bf 53} 9341 (1996).

\bibitem{hoh65} P. C. Hohenberg and P. C. Martin, Ann. Phys. (NY) {\bf
  34} 291 (1965).

\bibitem{gri93} A. Griffin {\it Excitations in a Bose-Condensed
  Liquid} (Cambridge: Cambridge University Press, (1993)).

\bibitem{lan80} L. D. Landau and E. M. Lifshitz {\it Statistical
  Physics, Third Edition, Part 1: Volume 5} (Butterworth-Heinemann, (1980)).

\bibitem{ket00} M. R. Andrews, D. M. Kurn, H. -J. Miesner,
  D. S. Durfee, C. G. Townsend, S. Inouye and W. Ketterle,
  Phys. Rev. Lett.  {\bf 79} 553 (1997).


\end{thebibliography}
\end{document}